\documentclass[%
 aip,
 jmp,%
 amsmath,amssymb,
 reprint,%
]{revtex4-1}

\usepackage{graphicx}% Include figure files
\usepackage{dcolumn}% Align table columns on decimal point
\usepackage{bm}% bold math
\begin{document}

\preprint{AIP/123-QED}

\title{Heavily $n$-doped Ge: low-temperature magnetoresistance properties}% Force line breaks with \\
%\thanks{Footnote to title of article.}

\author{A. Ferreira da Silva}
\affiliation{Instituto de F\'isica, Universidade Federal da Bahia,\\ 40210-340 Salvador, Bahia, Brazil.}%
%\email{Second.Author@institution.edu.}

\author{M. A. Toloza Sandoval}
\affiliation{Instituto de F\'isica, Universidade Federal da Bahia,\\ 40210-340 Salvador, Bahia, Brazil.}%
%\email{Second.Author@institution.edu.}

\author{A. Levine}%
\affiliation{Instituto de Fisica, Universidade de S\~{a}o Paulo
Laborat\'{o}rio de Novos Materiais Semicondutores,\\
05508-090 Butant\~{a}, S\~{a}o Paulo, Brazil}

\author{E. Levinson}
\affiliation{Instituto de Fisica, Universidade de S\~{a}o Paulo
Laborat\'{o}rio de Novos Materiais Semicondutores,\\
05508-090 Butant\~{a}, S\~{a}o Paulo, Brazil}

\author{H. Boudinov}
\affiliation{Instituto de Fisica, Universidade Federal do Rio
Grande do Sul,\\
91501 970  Porto Alegre, Rio Grande do Sul, Brazil}%

\author{B. E. Sernelius}
\affiliation{Division of Theory and Modeling, Department of
Physics, Chemistry
and Biology, Link\"{o}ping University, SE-581 83 Link\"{o}ping, Sweden}%


\begin{abstract}

Despite the groundbreaking provided by the development of the
heavily doped Ge and its applications, the understanding of its
fundamental properties remains incomplete. There are, in
particular, long-standing controversies regarding the conduction
mechanisms of such materials at low temperatures in magnetic
fields. We report here an experimental and theoretical study on
the magnetoresistance properties of the heavily phosphorous doped
germanium on the metallic side of the metal-nonmetal transition.
An anomalous regime, formed by negative values of the
magnetoresistance, was observed by performing low-temperature
measurements and explained within the generalized Drude model, due
to the many-body effects. It reveals a key mechanism behind the
magnetoresistance properties at low-temperatures and constitutes,
therefore, a path to its manipulation in such materials of great
interest of both fundamental physics and technological
applications.

\end{abstract}

%\keywords{Nanostructures | Spintronics}%Use showkeys class option if keyword
\maketitle

The advent of doped semiconductors stands for a milestone in the
development of the semiconductor devices. Since the seminals, with
p-n junction transistors\cite{Shockley} and solar
cells\cite{Chapin}, until the current trends, with mid-infrared
sensors and plasmonic devices\cite{Talie}, the doped
semiconductors have provided a fertile ground for fundamental
research and applied physics. Among the possibilities, such
materials can be used in energy-efficient windows \cite{SerBerJin,
HamGran}, because they can act as a metal for low photon energies
and as a semiconductor (or insulator) for high photon energies. In
comparison with ordinary metals, for which the carrier densities
are discrete and limited to a narrow range, doped semiconductors
constitute more flexible systems, allowing a continuous variation
of the carrier concentration over a wide range\cite{Shklovskii}.
In particular, such flexibility is even larger in $n$-doped
many-valley semiconductors, like Si and Ge. On the one hand, Si
has six equivalent band minima in the $\left\langle
{100}\right\rangle $-directions within the Brillouin zone (BZ); on
the other, Ge has eight in the L-points (the intersection of the
$\left\langle {111} \right\rangle $-directions with the zone
faces). Since the conduction-band valleys are strongly
anisotropic, when electrons are filling up the states at the
bottom of the conduction-band minima, they do not form Fermi
spheres, but cigar-shaped Fermi ellipsoids. As a consequence, the
contribution to transport and optical properties from each Fermi
volume is anisotropic. However, the sum of contributions from all
volumes is isotropic since the overall symmetry is cubic; note
that while the Si has six Fermi ellipsoids in the BZ, in the Ge
the eight minima lie at the BZ boundary, leading to electrons
effectively distributed within four Fermi ellipsoids.

It is well known that the application of uniaxial stress on the
sample breaks the afore-described symmetry, such that part of the
minima move upward in energy and part move downward, depending on
the applied stress direction. There is a redistribution of
electrons between the valleys and the applied stress results in
piezoresistance\cite{Fritz, SerPiezo} and optical
birefringence\cite{Feld, SerPiezo}. Furthermore, it is also
possible to modify the distribution of electrons by using an
external magnetic field. Each of the (aforementioned) Fermi
volumes is doubly degenerate and corresponds to spin-up and
spin-down electrons. With the introduction of a magnetic field,
the spin-up valleys move up in energy and the spin-down valleys
move down  - i.e., there is a redistribution of electrons where
the Fermi level is the same for both valley types. The system
remains isotropic, but important transport properties change, in
particular, the electric current parallel to the magnetic field,
which is expressed in terms of the longitudinal magnetoresistance,
as will be here discussed in detail.

\begin{figure}[t]
\includegraphics[width=8.0cm]{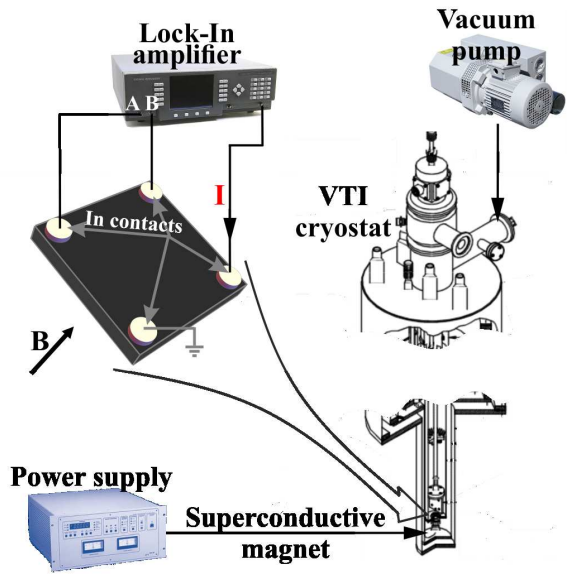}
\caption{Measurement setup consists of VTI (Variable Temperature
Insert) cryostat with the superconductive magnet, Lock-In
amplifier, and pump. GeP sample has 4 In contacts arranged in Van
der Paw geometry located at the corners of 7 mm$\times$7 mm
square. Magnetotransport measurements are done by the conventional
Lock-In technique with {\bf SIGNAL RECOVERY} (Model 7280) DSP dual
phase amplifier, which has a high input impedance of 100 M$\Omega
$. The sample was located in the superconductive magnet (Oxford)
with the perpendicular to its surface magnetic field up to 5
Tesla. Mechanical pump allowed us to reach temperatures down to
1.5 K} \label{figu2}
\end{figure}

For all conducting pure single crystals, the acquired knowledge
shows that, in general, the resistivity increases with the applied
magnetic field, i.e., the magnetoresistance is positive. On the
other hand, doped semiconductors require a detailed description at
the critical concentration, $n_c$, when the system turns metallic.
For densities much larger than $n_c$, if we place the donor
electrons at the bottom of the host conduction band and treat them
as a non-interacting electron gas, we found an unambiguous
agreement with experiments. However, an anomalous regime arises
when $n$ approaches $n_c$, in which, for example, the heat
capacity\cite{Kobay} and the spin susceptibility\cite{Quirt1,
Quirt2} are enhanced. In particular, low-temperature
magnetotransport properties are critically affected by this
regime, being the negative magnetoresistance a critical signature.
Theses so-called anomalies have attracted much attention with
several models reported in the literature\cite{Yam, Sas, Ion, And,
Zav, Emel, Ish, Kho, Toy, Alex}. With a peculiar interpretation,
Sernelius and Bergreen\cite{SerBer} proposed that the donor
electrons end up in the conduction band of the host already at the
critical concentration $n_c$ and suggested that the anomalous
properties, on the metallic side of and close to the transition
point, are caused by many-body effects\cite{Mahan}. One step
forward, we explore here such anomalous behavior of the
magnetoresistance in heavily n-doped Ge, comparing results from
low-temperature magnetotransport measurements with those obtained
from the theory.
As illustrated in Fig.\,\ref{figu2}, Hall and longitudinal
resistance measurements were performed in an Oxford cryostat with
VTI (Variable Temperature Insert), under a perpendicular magnetic
field provided by a superconducting coil. To prevent heating
effect and provide a clear signal for our measurements, was
employed the lock-in technique with frequencies 0.5-13 Hz in the
temperature range of  1.5-4.2 K  and bias current of 10 $\mu A$.
%%%%%%%%%%%%%%%%%%%%%%%%%%%%%%%%%%%%%%%%%%%%%%%%%%%%%%%%%%%%%%%%%%
\label{exp} %%%%%%%%%%%%%%%%%%%%%%%%%%%%%%%%%%%%%%%%%%%%%%%%%%%%

The samples were prepared in the following way: P-type, Ga doped
(100)-oriented square, $7 \times 7$\,mm$^2$, Ge samples with
resistivity in the range of 1-10 $\Omega {\rm{cm}}$ were implanted
with phosphorus at room temperature.
%%%%%%%%%%%%%%%%%%%%%%%%%%%%%%%%%%%%%%%%%%
\begin{figure}[t]
\includegraphics[width=7.5cm]{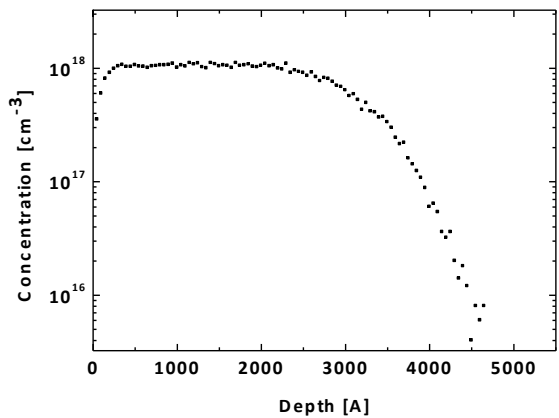}
\caption{Simulated multiple implantation phosphorous profile for
nominal sample atomic concentration of $10^{18} \rm{cm}^{-3}$.The
implanted P$^+$ doses were $2.0\times10^{13} \rm{cm}^{-2}$ (at 240
keV), $6.0\times10^{12} \rm{cm}^{-2}$ (at 140 keV),
$4.0\times10^{12} \rm{cm}^{-2}$ (at 80 keV), $2.0\times10^{12}
\rm{cm}^{-2}$ (at 40 keV), and $1.1\times10^{12} \rm{cm}^{-2}$ (at
20 keV).} \label{figu1}
\end{figure}
%%%%%%%%%%%%%%%%%%%%%%%%%%%%%%%%%%%
In each sample, implantations with energies of 240, 140, 80, 40,
and 20 keV were accumulated with appropriate doses to obtain a
plateau-like profile of P, from the surface to the depth of about
0.40 $\mu {\rm{m}}$, according to TRIM code
simulation\,\cite{Zieg}. In Fig.\,\ref{figu1} we show the
simulation for the concentration profile. To achieve a P atomic
concentration of $1 \times {10^{18}}\,{\rm{c}}{{\rm{m}}^{ - 3}}$,
the implanted P doses were $2.0 \times
{10^{13}}\,{\rm{c}}{{\rm{m}}^{ - 2}}$ (at 240 keV), $6.0 \times
{10^{12}}\,{\rm{c}}{{\rm{m}}^{ - 2}}$ (at 140 keV), $4.0 \times
{10^{12}}\,{\rm{c}}{{\rm{m}}^{ - 2}}$ (at 80 keV), $2.0 \times
{10^{12}}\,{\rm{c}}{{\rm{m}}^{ - 2}}$ (at 40 keV) and $1.1 \times
{10^{12}}\,{\rm{c}}{{\rm{m}}^{ - 2}}$ (at 20 keV).
The doses in the other samples were scaled to this sample,
according to the ratio of the desired P concentration.
Furthermore, the damage annealing and the electrical activation of
P were performed at 600 C for 1 minute in argon atmosphere in a
Rapid Thermal Annealing furnace to avoid high thermal budget; Van
der Pauw structures\cite{Pauw} were fabricated by applied indium
contacts at the corners of the samples and annealing at 80 C on a
hot plate for 1 minute was performed to improve the contacts. The
implantation process is described in
Refs.\,\cite{Ferr,Abram,Antonio}.

From the theoretical point of view, the conduction band of Ge has
four equivalent valleys ($\nu = 4$); there are eight minima in the
$\left( { \pm 1, \pm 1, \pm 1} \right)/\sqrt 3$ directions, but
they all are on the zone boundary so only half of each
cigar-shaped Fermi volume is inside the Brillouin zone. In heavily
$n$-type doped germanium, on the metallic side of the
metal-non-metal transition (i.e., $n > {n_c}$), the donor
electrons are up in the conduction band valleys. We consider that
the electrons are distributed in $\nu$ Fermi spheres and neglect
some known anisotropy effects on the resistivity\cite{Ser1}; the
relation between the radius of each sphere is then given
as\cite{Antonio} ${k_0} = {\left( {3{\pi ^2}n/\nu}\right)^{1/3}}$
and the Fermi energy given by ${E_0} = {\hbar^2}k_0^2/(2m{\rm{)}}
= {\hbar^2}k_0^2/\left( {2{m_{de}}{m_e}} \right)$, where ${{m_e}}$
is the electron rest mass and ${m_{de}} =0.22$ is the effective
mass of the density of states in one valley of the conduction
band. In particular, the contributions from the exchange and
correlation energy, ${E_{xc}}$, due to the influence of
ionized-donor potentials (the band-structure energy, ${E_b}$),
affect the parabolic band dispersion and the density of states.
Our model starts from the density of states from one valley, i.e.
\begin{equation}
\begin{array}{*{20}{l}}
{{D_E} = {D_k}/\left[ {dE\left( k \right)/dk} \right] =
\frac{{{k^2}}}{{{\pi ^2}\left[ {dE\left( k \right)/dk} \right]}},}
\end{array}
\label{equ1}
\end{equation}
and take into account that in each valley there are two states for
each ${\bf{k}}$ (i.e. one for each spin, up and down). Since
$D_E^0 = km/{\pi ^2}{\hbar ^2}$ is the density of states for
non-interacting electrons, the density of states for interacting
electrons can be expressed, in analogy, by introducing a
wave-number dependent effective mass, i.e.
\begin{equation}
{D_E} = k{m^*}/{\pi ^2}{\hbar ^2}, \label{equ3}
\end{equation}
with the effective mass given by
\begin{equation}
{m^*}\left( k \right) = m/\left[ {1 - \beta \left( k \right)}
\right], \label{equ4}
\end{equation}
where ${\beta}\left( k \right)$ gets a contribution from each of
the interaction energies, ${\beta }\left( k \right) =
\beta_{xc}\left( k \right) + \beta _b \left( k \right)$, such that
\begin{equation}
\begin{array}{l}
\beta _{xc} \left( k \right) =  - \frac{m}{{{\pi
^2}k}}\frac{\partial }{{\partial k}}\frac{{\delta N \cdot
{E_{xc}}}}{{\delta {n }\left( {\bf{k}} \right)}};\; \beta _b
\left( k \right) =  - \frac{m}{{{\pi ^2}k}}\frac{\partial
}{{\partial k}}\frac{{\delta N \cdot {E_b}}}{{\delta {n }\left(
{\bf{k}} \right)}}.
\end{array}
\label{equ5}
\end{equation}
$N$ is the total number of electrons and ${n }\left( {\bf{k}}
\right)$ is the occupation number of the state with wave-vector
${\bf{k}}$. Specially important for this paper, one effect of the
interactions is that around the Fermi level the effective mass and
density of states are enhanced\cite{Ser2}.

We use the generalized Drude model\cite{Ser3,Ser4} to calculate
the resistivity. For the static case, as here, the results agree
with the so-called Ziman's formula\cite{Ziman},
\begin{equation}
\begin{array}{l}
\rho = \frac{1}{{n{e^2}\tau /{m^*}}},\\
\frac{1}{\tau } = \frac{4}{3}\frac{{\nu {e^4}m}}{{\pi {\hbar
^3}{\kappa ^2}}}\int\limits_0^{2k_0} {dq\frac{1}{{q{{\tilde
\varepsilon }^2}\left( {q,0} \right)}}},
\end{array}
\label{equ6}
\end{equation}
where $\rho $, $\tau $ and $\kappa$ are respectively the
resistivity, transport time and dielectric constant
($\kappa=15.36$ for Ge).

The presence of a static and spatially homogeneous magnetic field
{\bf B} leads to a redistribution of electrons between spin up and
spin down bands, which affects the density of states, the
effective mass at the Fermi level, the conductivity and the
transport time. Let us introduce the spin-polarization parameter,
$s$, that varies from zero in absence of {\bf B} to $1$ at full
polarization (all electrons have spin down),
\begin{equation}
s = \frac{{n ^{\downarrow} - n ^{\uparrow} }}{n}. \label{equ7}
\end{equation}

For spin up and down electrons, the density and Fermi wave-number
are respectively,
\begin{equation}
\begin{array}{*{20}{lcl}}
{n^ \uparrow }& = & n(1 - s)/2,\\
{n^ \downarrow }& = & n(1 + s)/2,\\
{k_0}^ \uparrow & = & {k_0}{{\left( {1 - s} \right)}^{1/3}},\\
{k_0}^ \downarrow & = & {k_0}{{\left( {1 + s} \right)}^{1/3}.}
\end{array}
\label{equ8}
\end{equation}

Therefore, the resistivity is now written as\cite{Antonio}
\begin{equation}
\rho \left( s \right) = \frac{{m/{e^2}}}{{{n^ \uparrow }{\tau ^
\uparrow }\left( {1 - {\beta ^ \uparrow }} \right) + {n^
\downarrow }{\tau ^ \downarrow }\left( {1 - {\beta ^ \downarrow }}
\right)}}. \label{equ9}
\end{equation}
%%%%%%%%%%%%%%%%%%%%%%%%%%%%%%%%%%%%%%%%%%%%%
\begin{figure}[t]
\includegraphics[width=7.5cm]{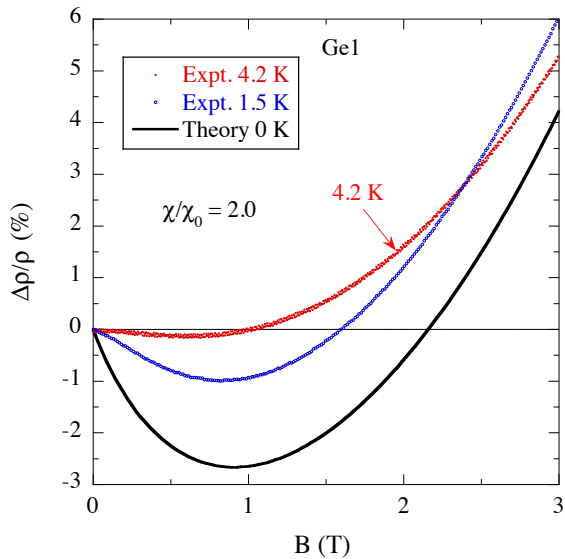}
\caption{(Color online) The magnetoresistance at the temperatures
4.2 K (red curve) and 1.5 K (blue curve) as function of magnetic
induction, B, for a Ge:P sample with doping concentration
$2.96\times10^{17}$ cm$^{-3}$.  The black solid curve is our
theoretical result for 0 K. See the text for details.}
\label{figu3}
\end{figure}

Note that the magnetoresistance is given by $\Delta \rho /\rho  =
\left[ {\rho \left( s \right) - \rho \left( 0 \right)}
\right]/\rho \left( 0 \right)$, i.e. it is a function of the spin
polarization $s$; however, the experimental results are given in
terms of $B$. When the modulus of the magnetic field is small
enough, one can assume the following linear relation between $B$
and $s$:
\begin{equation}
B\left[ T \right] = \frac{{2.64262 \times {{10}^{ - 11}}{{\left(
{n\left[ {c{m^{ - 3}}} \right]/\nu }
\right)}^{2/3}}}}{{{m_{de}}\left( {\chi /{\chi _0}} \right)}}s.
\label{equ10}
\end{equation}
%%%%%%%%%%%%%%%%%%%%%%%%%%%%%%%%%%%%%%%%%%%%%%%

\begin{figure}[t]
\includegraphics[width=7.5cm]{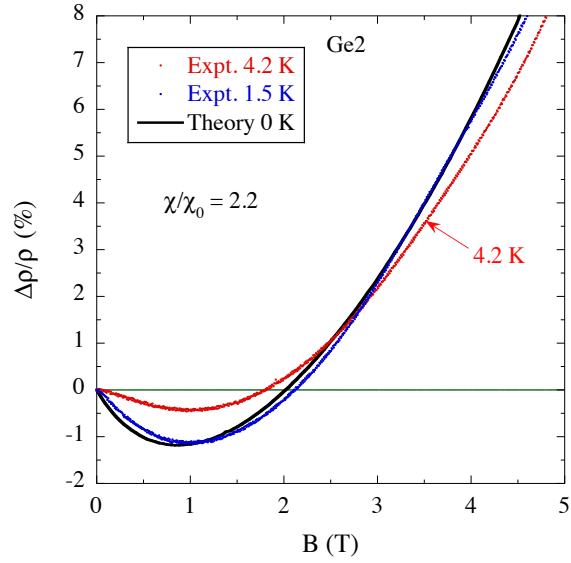}
\caption{(Color online) The same as Fig.\,\ref{figu3}, but here
for the doping concentration $6.25\times10^{17}$ cm$^{-3}$. }
\label{figu4}
\end{figure}

\begin{figure}[t]
\includegraphics[width=7.5cm]{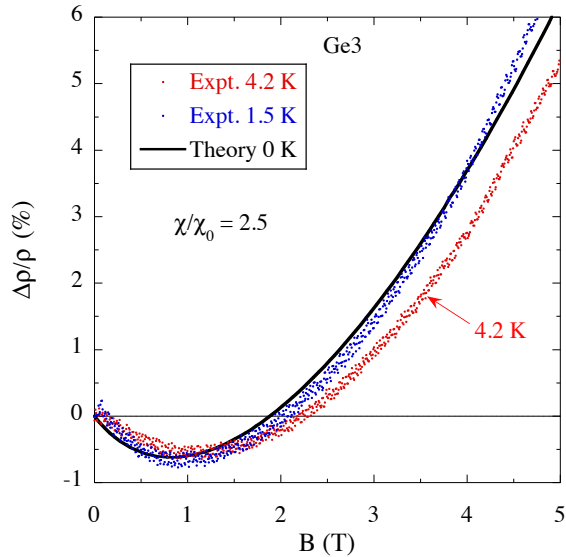}
\caption{(Color online) The same as Fig.\,\ref{figu3}, but here
for the doping concentration $1.17\times10^{18}$ cm$^{-3}$.}
\label{figu5}
\end{figure}

We compare obtained theoretical and experimental results in
Figs.\,\ref{figu3} - \ref{figu5}. The spin-susceptibility
enhancement-factor (${\chi /{\chi_0}}$) and effective mass
($m_{de}$) were adjusted\cite{note2} to optimize the fit between
theoretical and experimental curves; note, however, that this
adjustment does not affect our main picture, with negative values
for the magnetoresistivity as well as its signal inversion. In
Fig.\,\ref{figu3} we show the results for the sample with the
lowest doping concentration, which is closest to the
metal-nonmetal transition (reminding that\cite{Mott} $n_c\approx
2.5\times10^{17}\rm{cm}^{-3}$) and for which the magnetoresistance
presents a minimum that becomes deeper when the temperature
decreases. The black line shows the theoretical curve obtained for
the spin-susceptibility enhancement-factor equal to 2 (and 0 K),
and the blue and red lines correspond respectively to experimental
results for 1.5 K  and 4.2 K. Fig.\,\ref{figu4} presents the
results for the sample with the next lowest doping concentration;
in comparison with the Fig.\,\ref{figu3}, we see a more shallow
minimum for the theoretical curve and little deeper minima for the
experimental curves. Here, we use  ${\chi /{\chi _0}}$ equal to
2.2.

\begin{figure}[b]
\includegraphics[width=7.5cm]{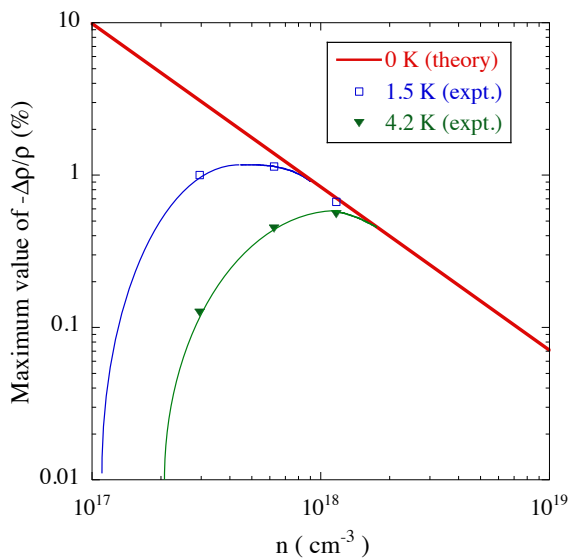}
\caption{(Color online) The depth of the magnetoresistance minima
as function of doping concentration. The $red$ thick solid
straight line is the theoretical result for 0 K; the $blue$ open
squares are the experimental results at 4.2 K; the $green$ filled
triangles are the experimental results for 1.5 K; the thin solid
curves are just guides for the eye. See the text for details.}
\label{figu6}
\end{figure}

In Fig.\,\ref{figu5} we present the results for the sample with
the highest doping concentration, where we use ${\chi /{\chi _0}}$
equal to 2.5. Analyzing the Figs.\ref{figu3} - \ref{figu5}, we
identify two competing effects: while the lowering of the doping
density leads to deeper minima, the increment of the temperature
leads to shallower minima. It is also important to note that, near
and on the metallic side of the metal-nonmetal transition, the
enhancement of the density of states at the Fermi level increases.
Furthermore, the enhancement of the spin susceptibility also
increases when the density comes closer to $n_c$; however, it is
reduced when the temperature goes
up\cite{Quirt1,Quirt3,Quirt2,Ferr2}.
%%%%%%%%%%%%%%%%%%%%%%%%%%%%%%%%%%%%%%%
%%%%%%%%%%%%%%%%%%%%%%%%%%%%%%%%%%%%%%%
Using a log-log plot, in Fig.\ref{figu6} we show how the maximum
negative magnetoresistance decreases linearly when the doping
concentration increases. Note that the maximum starts to decrease
at a density that depends on the temperature. The higher the
temperature, earlier the maximum starts to decrease, as also
reported in Ref.\cite{Yam}.

\begin{figure}[t]
\includegraphics[width=7.5cm]{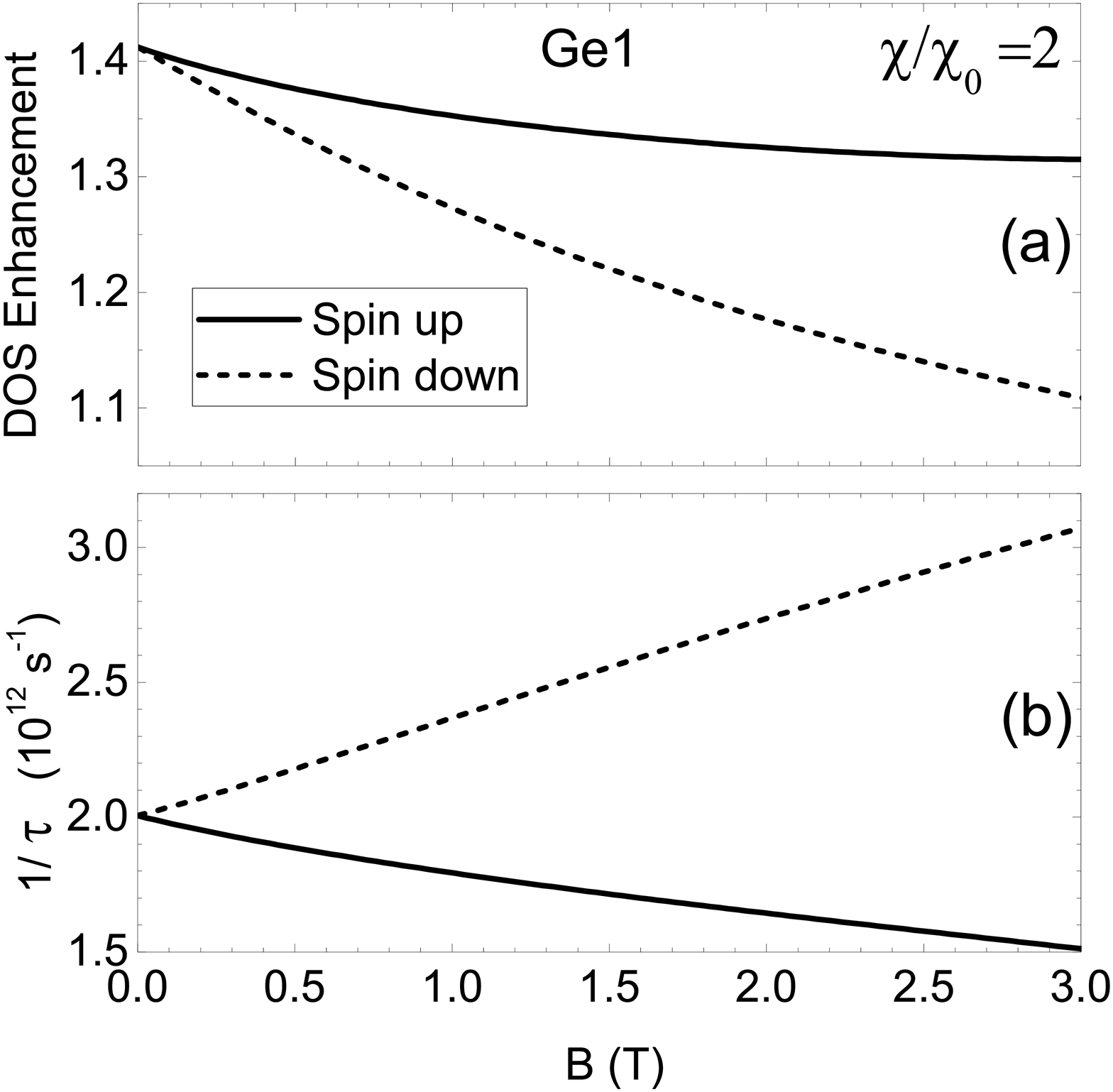}
\caption{(a) The enhancement of the density of states at the Fermi
level for spin up and spin down electrons as functions of the
magnetic induction B. (b) The inverse transport time for spin up
and spin down electrons as functions of B. In (a) and (b) we
consider the doping concentration $2.96 \times10^{17}$ cm$^{-3}$.}
\label{figu7}
\end{figure}

Here we propose an explanation to the cause of the negative
magnetoresistance observed at low temperatures in heavily
phosphorous doped germanium on the metallic side of the
metal-nonmetal transition. First, in the absence of magnetic
fields, the density-of-states enhancement at the Fermi level
contributes to the enhanced resistivity. Second, the presence of a
magnetic field lifts the degeneracy of the electron dispersion,
resulting in an upshifted spin-up band and a downshifted spin-down
band. At the Fermi level, there is a redistribution of electrons
between spin-up and spin-down bands, which leads spin-up and
spin-down electrons to states with wave-numbers ${k_0} ^{\uparrow}
$ and ${k_0}^{\downarrow} $ respectively. Consequently, the peak
corresponding to the density of states at the Fermi-level splits
in ${k_0} ^{\uparrow} $ and ${k_0}^{\downarrow} $: for electrons
with ${k_0} ^{\uparrow} $, one peak remains at the Fermi-level
while the other moves down into the unoccupied part of the bands;
instead, for electrons with ${k_0}^{\downarrow} $, while one peak
remains at the Fermi-level, the other moves up into the occupied
part of the bands. The enhancement at the Fermi-level is, then,
reduced for both spin types.
In Fig.\ref{figu7}(a) we show the enhancement of the density of
states at the Fermi-level, for both spin up and spin down, as
functions of the magnetic-field modulus, considering the lowest
doping concentration (i.e. $2.96\times10^{17}$ cm$^{-3}$). We are
also considering that only the enhancement at the Fermi level
affects the resistivity and that the effect due to the scattering
against Friedel oscillations\cite{Friedel}, which eventually
contributes to the enhancement of the resistivity\cite{Antonio},
can be negligible in heavily $n$-doped Ge. For completeness, we
show in Fig.\ref{figu7}(b) how the scattering rates for spin-up
and spin-down electrons vary with the magnetic-field modulus.

Our model considers the temperature equal to zero, but the
knowledge acquired from experiments shows that the
magnetoresistance reduces when the temperature
increases\cite{Yam}. To interpret this well- known behavior, note
that the peak of the density of states at the Fermi-level is
expected to be broadened and only states at the Fermi-level
contribute to the conductivity, at zero temperature. The
temperature effect enables states away from the Fermi-level, for
which enhancement of the density of states is weaker, to
contribute to the conductivity, and we expect that these effects
gradually remove the negative magnetoresistance. Furthermore, the
temperature effects become more important for lower densities, as
can be seen in our experimental results as well as in
Ref.\cite{Yam}.

%%%%%%%%%%%%%%%%%%%%%%%%%%%%%%%%%%%%%%%%%%%%%%%%%%%%%%%%%%%%%%%%%%%

To summarize and conclude, we have investigated the anomalous
regime of the longitudinal magnetoresistance of heavily $n$-doped
germanium on the metallic side of the metal-non-metal transition,
by using magnetotransport measurements at low temperatures (1.5 K
and 4.2 K) and comparing with obtained results from many-body
theory, where the donor-electrons are assumed to reside at the
bottom of the many-valley conduction band of the host.
For doping densities above and close to $n_c$, we found a regime
formed by negative values of the magnetoresistance that is
drastically suppressed when the temperature increases and
physically interpreted in terms of many-body effects. The obtained
results show that the experiments support the model and can help
in understanding the mechanism of magnetoresistance of heavily
doped semiconductors. Additionally, more samples in the doping
range of $10^{18}-10^{19} \rm{cm}^{-3}$ would be helpful for
further verification of the theory.

The authors acknowledge financial support from the Brazilian
agencies: CNPq (Proj. 303304/2010-3), CAPES (PNPD
88882.306206/2018-01), FAPESB (PNX 0007/2011 and INT 0003/2015)
and FAPESP (Proj. 15/16191-5).

\nocite{*}
%\bibliography{aipsamp}% Produces the bibliography via BibTeX.

\begin{thebibliography}{[1]}
\bibitem{note2} We have enhanced the effective mass, $m_{de}$,
characterizing the Fermi spheres with a factor of 1.5, i.e. it has
been changed from .22 to .33. Also note that the adjustment of
${\chi /{\chi_0}}$ only affects the theoretical curves in the
horizontal direction.
\end{thebibliography}

\end{document}